# LOFT - Large Observatory for X-ray Timing

**S. Zane**[a][*], **on Behalf of the LOFT Detector's Group**

[a]*Mullard Space Science Laboratory, University College London,
Holmbury St Mary, Dorking, Surrey, RH56NT, UK*
*E-mail:* `s.zane@ucl.ac.uk`

ABSTRACT: LOFT (the Large Observatory for X-ray Timing), is a mission concept that was considered by ESA as a candidate for an M3 mission and has been studied during an extended >2-years long assessment phase.
The mission was specifically designed to perform fast X-ray timing and probe the status of the matter near black holes and neutron stars. The LOFT scientific payload is composed of a Large Area Detector (LAD) and a Wide Field Monitor (WFM). The LAD is a 10 $m^2$-class pointed instrument with ∼15 times the collecting area of the largest past timing missions (as the Rossi XTE) over the 2-30 keV range (30-80 keV expanded), combined with CCD-class spectral resolution, which holds the capability to revolutionise studies of X-ray variability down to the millisecond time scales.
Its ground-breaking characteristic is a mass per unit surface in the range of ∼10 kg/$m^2$, enabling an effective area of ∼10 $m^2$ (at 10 keV) at a reasonable weight. The development of such large but light experiment, with low mass and power per unit area, is now made possible by the recent advancements in the field of large-area silicon drift detectors and capillary-plate X-ray collimators. Although the LOFT mission has not been down-selected for launch in the M3 ESA programme (with launch in 2022-2024), during the assessment phase most of the trade off have been closed leading to a robust and well documented design which will be re-proposed in the future ESA calls. In this paper, we will summarize the characteristics of the LAD instrument and briefly describe the status of the detectors design.

---

[*]Corresponding author.

# Contents

## 1. Introduction

LOFT (the Large Observatory for X-ray timing) is a mission design that has been studied by ESA as a candidate for being an M3 mission within the Cosmic Vision programme (see ref. [1, 2, 3, 4] and references therein). The mission is devoted to high resolution X-ray timing, and specifically designed to investigate the space-time around collapsed objects and the matter inside them.

The rationale behind the mission concept is quite simple. Since the beginning of X-ray timing astronomy, the number of new discoveries and their scientific impact has been in direct correlation with the collecting area of the instrument. Explorations started with Uhuru (which had an area of 840 cm$^2$), then proceeded with EXOSAT (1600 cm$^2$), Ginga (4000 cm$^2$), and culminated with the RoSSI XTE, that, with an area of 6500 cm$^2$, has been the larger observatory launched until now and left the larger heritage. The main challenge of LOFT is to make a major leap forward in this direction, reaching a size ∼15 times larger than the Rossi XTE, and in turn by revolutionizing our knowledge of X-ray variability. Specifically, LOFT will assess three main scientific goals with the science theme 3 proposed in the ESA Cosmic Vision programme: "What are the fundamental physical laws of the Universe". First, it aims to probe the behavior and motion of matter in the presence of strong gravitational fields (in the stationary spacetime of black holes and neutron stars down to a few gravitational radii). where the strong field effects predicted by General Relativity are largest. Second, through a series of neutron star observations it will make a robust diagnostic of the physics of matter at supra nuclear densities, determining its equation of state and composition and in turn decoding the strong force nature. Third, it will perform spectral and timing observations for virtually all classes of X-ray transients with an unprecedented throughput.

The LOFT mission concept was studied by a consortium of European scientific institutes, including teams from the Czech Republic, Denmark, Finland, France, Germany, Italy, the Netherlands, Poland, Spain, Switzerland and the United Kingdom, with support from international partners in Brazil, India, Japan and the United States. An even wider science support community contributed to the science case.

## 2. The LOFT payload

As mentioned in the previous section, LOFT is intended to answer fundamental questions about the motion of matter orbiting close to the event horizon of a black hole, and the state of matter in neutron stars. In practice, these scientific objectives will be achieved through high-resolution X-ray timing, i.e. through the measurement of X-ray photometric time series and spectra from a range of astrophysical compact objects (neutron stars, Galactic black holes and extra-Galactic black holes in AGNs). These measurements require no imaging capability, but instead a 10 m$^2$ class instrument, in combination with good spectral resolution. The LOFT payload comprises two instruments (see figure 1). The main pointed timing instrument is the LAD (see ref. [3]), made of 6 deployable panels (in green in figure 1) tiled with Silicon Drift Detectors (SDD) and Microchannel Plates (MCP). The panels have a total geometrical area of 18 m$^2$ (collimated area of $\sim$ 10 m$^2$). The instrument operates in the energy range 2-30 keV (up to 80 keV in expanded mode) with a time resolution of 10 $\mu$s and an energy resolution of $\sim$260 eV at 6 keV. The 1-deg collimated field of view LAD will be able to access $\sim$ 75% of the sky at any time, to observe Galactic and bright extragalactic X-ray transients at different flux levels. The second instrument, the WFM (in yellow in figure 1, see ref. [4]), is a coded mask detector that operates in the same energy range of the LAD with a position accuracy of 1 arc-minute. The WFM will monitor more than half of the LAD-accessible sky (approximately 1/3 of the whole sky) simultaneously at any time and in the same energy range, providing information about source state (flux variability and energy spectrum) and discovering new transients. In addition, LOFT will host on board a burst alert system which will allow to trigger bright transient events (such us GRBs, thermonuclear X-ray bursts, etc), to download their position within few tens of seconds from the event and timing or spectral information within 2-3 hours .

The key to the 20x breakthrough in effective area that can be achieved by the LAD resides in the synergy between technologies imported from other fields of scientific research, both ground- and space-based. In particular, the crucial ingredients for a sensitive but lightweight experiment, enabling $\sim$ 18 m$^2$ geometric area payload at reasonable weight, are the innovative large-area SDDs designed at INFN Trieste (see next section). The unambiguous identification of the target source is then reached by narrowing the field of view by means of an aperture collimator, down to a level of $\sim$ 1 deg, large enough to allow for pointing uncertainties yet small enough to reduce the aperture background and the risk of source confusion. The LAD is therefore designed as a classical collimated experiment. The lightweight glass collimator is based on the technology of the capillary-plates (the same as the standard microchannel plates) which has already been applied within the MEDA and GSPC instruments onboard the EXOSAT mission as well as currently baselined for the MIXS experiment onboard the ESA BepiColombo mission (see ref. [5]).

The basic LAD detection element (detector) is composed of a SDD, its front-end electronics (FEE), and a collimator. The 6 detector panels will be tiled with 2016 detectors, electrically and mechanically organized in groups of 16, referred to as modules. Each of the 6 panels hosts 21 modules. The assembly philosophy employs a hierarchical approach: detector, module, detector panel and LAD Assembly (see figure 1). The FEEs of the 16 detectors in a module converge into a single Module Back End Electronics (MBEE). Finally, the 21 MBEE in each panel converges into a Panel Back-End Electronics (PBEE) which is in charge of interfacing them in parallel, making

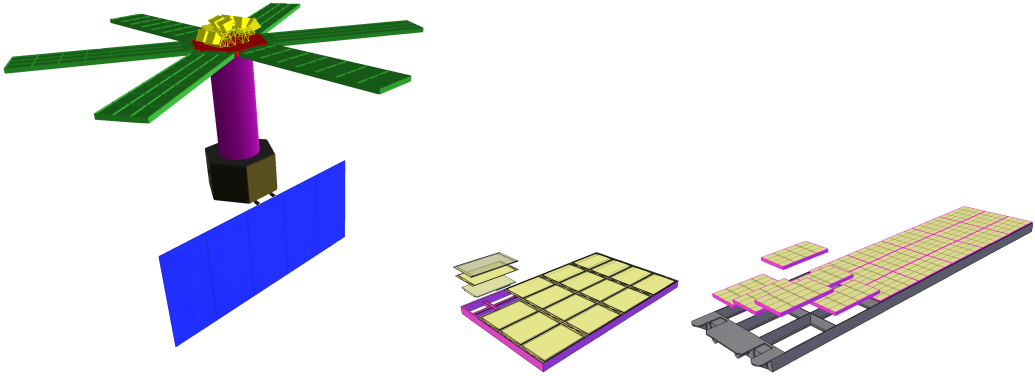

**Figure 1.** Left: The configuration for LOFT as baselined by the LOFT Consortium in the M3 proposal. Green = LAD, yellow = WFM, Red = Optical bench, Purple = Structural Tower, Gold = Bus, Blue = Solar array. Center: Front view of a module, showing the tiling on the SDD, FEE and collimator. Right: Each LAD panel is populated with 21 modules.

the module the basic redundant unit from the mechanical and read-out point of view.

## 3. The LOFT detectors

The LAD and WFM detectors share the same technology and are based on the heritage of the detectors used in the Inner Tracking System of the ALICE experiment at the Large Hadron Collider at CERN (see ref. [6, 7, 8]), for which INFN Trieste had and maintains the full knowledge of both design and process details.

The working principle of the SDD is summarized in figure 2 (left panel): the charge generated by the absorption of an X-ray photon is collected in the middle plane of the detector and then it is drifted towards the read-out anodes on one edge of the detector. A series of cathodes on both sides of the detector maintain the electric field; the operation of the LOFT SDD requires a 1300V bias. Table 2 summarizes the main characteristics of the LAD and WFM SDDs. For comparison, the ALICE detectors have a thickness of 300 $\mu$m, an active area of 53 cm$^2$, a drift time of 5 $\mu$s, an anode pitch of 294 $\mu$m and a single channel area of 0.1 cm$^2$. In order to meet the LOFT requirements, the ALICE design must be modified and the most challenging goals are the increase in area and pitch angle.

During the >2 years of LOFT assessment phase we have produced a number of prototypes converging to the final design. All LOFT/LAD and WFM prototype detectors have been manifactured at FBK (Fondazione Bruno Kessler, Trento, Italy), and the design was lead by INFN, Trieste (see ref. [9, 10]). The LAD sensors feature the following characteristics: 2-50 keV energy range with a spectral resolution below 240 eV at 6 keV, power consumption below 0.5 mW/cm$^2$ of sensitive area at room temperature, enhanced quantum efficiency at the low end of the energy range ($\sim$ 39% at 2 keV, maximum of 97% at 9 keV), and the possibility of X-ray spectroscopy timing on a ten microseconds scale.

The final LAD prototype, delivered by the end of 2013, was produced using 6-inch diameter floating-zone Silicon wafers, with a resistivity of approximately 9 k$\Omega\times$cm and a thickness of 450 $\mu$m (see figure 2, right panel). The silicon tile containing the SDD (sensitive area of about 108 $\times$ 70 mm$^2$) is a rectangle with an area of approximately 120 $\times$ 72 mm$^2$, the largest cut out which is possible to obtain from a 6-inch diameter wafer without changing the drift length. The detector is symmetrical with respect to the central p$^+$ cathode: electrons drift from the central p$^+$ cathode towards two linear arrays of readout n$^+$ anodes positioned at the two sides. For each half-detector there are 292 drift cathodes with a pitch of 120 $\mu$m and 112 readout anodes with a pitch of 970 $\mu$m. The sensitive-to-total-area ratio is 87%. The detector, which is the largest SDD ever produced, is planned to work at a drift field of 360 V/cm (HV bias of $\sim$1300 V), entailing a maximum drift time of about 7 $\mu$s at 20 $^\circ$C, which is reduced to $\sim$5 $\mu$s at -20 $^\circ$C due to the higher electron mobility at lower temperatures. We refer the interested reader to [10] for all details.

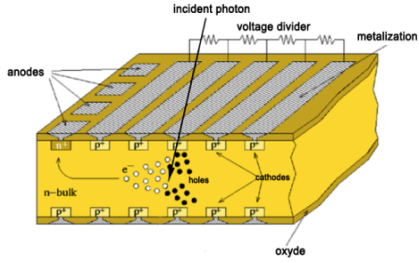
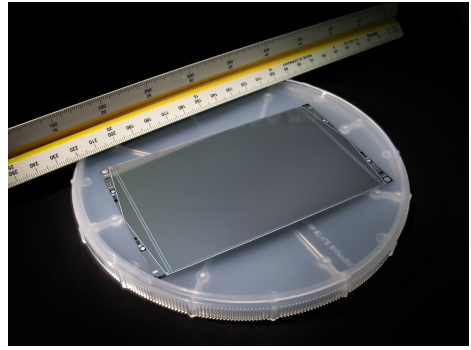

**Figure 2.** Left: The working principle of the SDD. Right: the 6-inch prototype for the LOFT LAD detector (see text for details and references).

**Table 1.** The LOFT LAD and WFM SDDs main parameters

| Parameter | LAD | WFM |
| --- | --- | --- |
| Thickness | 450 $\mu$m | 450 $\mu$m |
| Geometric Size | 120.3 mm $\times$ 72.5 mm | 77.4 mm $\times$ 72.5 mm |
| Active Area | 108.5 mm $\times$ 70.0 mm | 65.1 mm $\times$ 70.0 mm |
| Anode Pitch | 970 $\mu$m | 145 $\mu$m |
| Drift Length | 35 mm | 35 mm |
| Drift Field | 360 V/cm | 360 V/cm |
| Number of Anodes | 2 $\times$ 112 | 2 $\times$ 448 |
| Anode Capacitance | 350 fF | 81 fF |

## 4. Radiation damage and tests

Space weather has a number of effect on SDD that results in a degradation of the SDD spectral resolution by radiation damage in orbit. The most important ones are: a) the Total Ionising Dose (TID), aka a measure of the energy absorbed by matter (computed in rad, radiation absorbed dose, such us 100rad=1 J/kg. In a Low Earth Orbit, LEO, the TID is mostly due to e$^-$ and protons trapped in the inner belt); b) the Non Ionizing Energy Loss (NIEL), i.e. a displacement damage due to cumulative long-term non-ionizing damage from protons, electrons and neutrons; and c) a Space radiation effect on the Charge Collection Efficiency (CCE) of detectors. In order to quantify such effects, we carried out a careful and in-depth study of the radiation environment. Radiation effects on SDDs, mounted with an ASIC (see ref. [11]) have been verified by means of several (six) irradiation test campaigns (see ref. [12, 13] and the presentation of Y. Evangelista at this meeting):

- TID, NIEL, CCE:

    1. 1 GeV electrons irradiation tests (ALICE prototype, 2002);
    2. 50 MeV proton irradiation tests (ALICE prototype, 2011);
    3. 10 MeV proton irradiation tests (XDXL2 prototype, 2013);
    4. 0.2-0.8 MeV proton irradiation tests (XDXL1-2 prototype, 2012-2013);

- Debris/Micro-meteorites:

    1. 0.5-2.5 $\mu$m diameter, 0.5-12 km/s speed irradiation tests (XDXL2 prototype, 2012-2013);

- Environmental:

    1. Thermal: routine operation in -35°C / +30 °C;
    2. Vacuum: switch-on/off and long operation at $2 \times 10^{-6}$ mbar.

We have found that, in the case of LOFT, TID is negligible (<2 krad at the end of life, EOL, of the mission), CCE negligible (< 0.1%), while NIEL is the main effect. In order to mitigate this, we suggest a combination of the orbit selection, which provides best protection, with a setting of the SDD operating temperature requirement. Specifically, we decided to operate the detectors at a temperature below -10 °C (for the nominal orbit with 550 km altitude and 0° inclination). The operative temperature is reached by means of a "passive" temperature control system, using radiators in both the LAD and WFM. Results of the debris/micro-meteorites bombardment test have shown that an impact produces a sudden increase of the device leakage current (of order of 1-200 pA), confined to the crater region. While in the LOFT design the SDDs of the LAD are protected by the lead glass collimator, with a field of view of $3 \times 10^{-4}$sr, and the expected particle rate is only $\sim 7 \times 10^{-2}$ per year for the whole LAD, the detectors on the WFM require a dedicated shielding, a 25 $\mu$m thick beryllium located above the detection plane.

## 5. Summary

A significant activity has been carried out by INFN Trieste and FBK to design, prototype and test innovative large area SDDs, as required by the LOFT mission. The LOFT mission concept has been studied by the LOFT Consortium, ESA and two potential industrial primes for more than 2 years leading to a unique and well documented science case supported by a robust payload design, which will be reproposed in the future ESA calls.

The LOFT detector's groups consists of: M. Ahangarianabhari, P. Azzarello, D. Barret, G. Bertuccio, E. Bozzo, C. Budtz-Jorgensen, F.R. Cadoux, R. Campana, E. Del Monte, J.W. den Herder, Y.

Evangelista, S. Fabiani, Y. Favre, M. Feroci, G.W. Fraser, F. Fuschino, D. Haas, R. Hudec, C. Labanti, F. Lebrun, P. Malcovati, M. Marisaldi, V. Petracek, M. Pohl, I. Rashevskaya, M. Rapisarda, A. Rachevski, A. Vacchi, G. Zampa, N. Zampa.

**Acknowledgments**

The work of the MSSL-UCL and Leicester groups is supported by the UK Space Agency. The Italian team is grateful for support by ASI (under contract I/021/12/0), INAF and INFN. The work of SRON is funded by the Dutch national science foundation (NWO). The work of the group at the University of Geneva is supported by the Swiss Space Office. The work of IAAT on LOFT is supported by Germany's national research center for aeronautics and space - DRL. The work of the IRAP group is supported by the French Space Agency. LOFT work at ECAP is supported by DLR under grant number 50 00 1111. RH (CTU in Prague) acknowledges GACR grant 13-33324S.